\documentclass[aps,prl,twocolumn,superscriptaddress,longbibliography]{revtex4-2}
\usepackage{graphicx}
\usepackage{dcolumn}
\usepackage{bm}
\usepackage{color}
\usepackage[colorlinks=true,allcolors=blue,breaklinks=true]{hyperref}
\usepackage{amsmath}
\usepackage[normalem]{ulem}

\begin{document}

\title{Experimental validation of a micromechanically-based compaction law for soft/hard grain mixtures}

\author{Manuel C\'{a}rdenas-Barrantes}
\email{manuel-antonio.cardenas-barrantes@umontpellier.fr}
\affiliation{Laboratoire de M\'{e}canique et G\'{e}nie Civil, UMR 5508 CNRS-University Montpellier, 34095 Montpellier, France}
\author{Jonathan Bar\'{e}s}
\email{jonathan.bares@umontpellier.fr}
\affiliation{Laboratoire de M\'{e}canique et G\'{e}nie Civil, UMR 5508 CNRS-University Montpellier, 34095 Montpellier, France}
\author{Mathieu Renouf}
\email{Mathieu.Renouf@umontpellier.fr}
\affiliation{Laboratoire de M\'{e}canique et G\'{e}nie Civil, UMR 5508 CNRS-University Montpellier, 34095 Montpellier, France}
\author{Émilien Az\'{e}ma}
\email{emilien.azema@umontpellier.fr}
\affiliation{Laboratoire de M\'{e}canique et G\'{e}nie Civil, UMR 5508 CNRS-University Montpellier, 34095 Montpellier, France}
\affiliation{Institut Universitaire de France (IUF), Paris, France}

\date{\today}

\begin{abstract}
In this letter, we report on an experimental study which analyzes the compressive behavior of 2D bidisperse granular assemblies made of soft (hyperelastic) and hard grains in varying proportions ($\kappa$).
By means of a recently developed uniaxial compression set-up \cite{vu2019_pre} and using advanced Digital Image Correlation (DIC) method, we follow, beyond the jamming point, the evolution of the main mechanical observables, from the global scale down to the strain field inside each deformable grain. First, we experimentally validate and extend to the uni-axial case a recently proposed micro-mechanical compaction model linking the evolution of the applied pressure $P$ to the packing fraction $\phi$ \cite{cantor2020_prl}. Second, we reveal two different linear regimes depending on whether the system is above or below a cross-over strain unraveling a transition from a discrete to a continuous-like system. Third, the evolution of these linear laws are found to vary linearly with $\kappa$, up to a saturation point around $\kappa=80$\% of hard particles. These results provide a comprehensive experimental and theoretical framework that can now be extended to a more general class of polydisperse soft granular systems.
\end{abstract}

\maketitle


Granular materials are ubiquitous in nature and human activities. In their most generic form, they are composed of grains of shape, dimension and more importantly bulk properties that are very different, even in the same packing. Among them, granular systems composed of squeezable compounds are the ones with the most singular behaviors. This is the case of biological tissues composed of soft cells \cite{wyatt2015_pnas,mauer2018_prl}, liquid foams \cite{bolton1990_prl,katgert2010_epl}, emulsions \cite{brujic2003_fd,zhou2006_sci}, or sintered material \cite{cooper1962_jacs,kawakita1971_pt} to name a few. The specificity of materials composed of both soft and hard particles lies in the fact that under any loading, and even deep in the jammed state \cite{liu1998_nat}, two mechanisms compete in the system: particle rearrangement and deformation. When loaded, hard particles have only the ability to rearrange abruptly \cite{hayman2011_pag,bares2017_pre} while it is easier for soft particles to deform in order to sustain a given strain \cite{vu2019_pre}. In these materials, both mechanisms cohabit, prioritizing either one or the other depending on the soft to hard particle ratio and global loading.

Compressed systems made of purely hard or slightly deformable particles have been extensively studied these last decades from a theoretical \cite{liu1998_nat,jacquin2011_prl,van2009_jpcm}, numerical \cite{radjai1996_prl,ohern2003_pre} and experimental \cite{majmudar2007_prl,cox2016_epl,papadopoulos2016_pre} point of view. Most of the fundamental aspects of their behavior under many solicitations are now well-known \cite{liu1998_nat,jop2006_nat,van2009_jpcm}. The behavior of systems made of highly deformable grains at high packing fraction is also more and more studied \cite{zhou2006_sci,vu2019_pre,brujic2003_fd,cardenas2020_pre,cardenas2021_pre} and at least in compaction, reliable models exist \cite{cantor2020_prl}. In between, and even if they are very common materials, systems composed of grain with very different -- soft and hard -- rheologies, driven far above the jamming transition, have a bewildering but fascinating behavior that stays mostly misunderstood. They have been experimentally studied in very specific applications like for stress releasing \cite{indraratna2019_geo,khatami2020_gi}, seismic isolation \cite{tsang2008_ehsd,tsiavos2019_sdee} or foundation damping \cite{anastasiadis2012_gge,mashiri2015_sf}. However, to our best knowledge, no local measurements have been performed to understand micro-processes leading to their very specific macroscopic behavior. Only very recently, through numerical approach, a micro-mechanical-based compaction model has been proposed \cite{cardenas2020_pre,cardenas2021_pre} to describe the evolution of these systems in compression. But an experimental validation is still lacking as well as a clear understanding of local processes.

In this Letter, we aim to fill this gap by extending the recently proposed compaction law and experimentally validating it, and by stating constitutive laws linking both local structural parameters and local strain field with global observables such as pressure and packing fraction. The evolution of these laws as a function of the fraction of soft grain is also investigated, and saturation limits observed.


\begin{figure}
\centering
\includegraphics[width=0.90\linewidth]{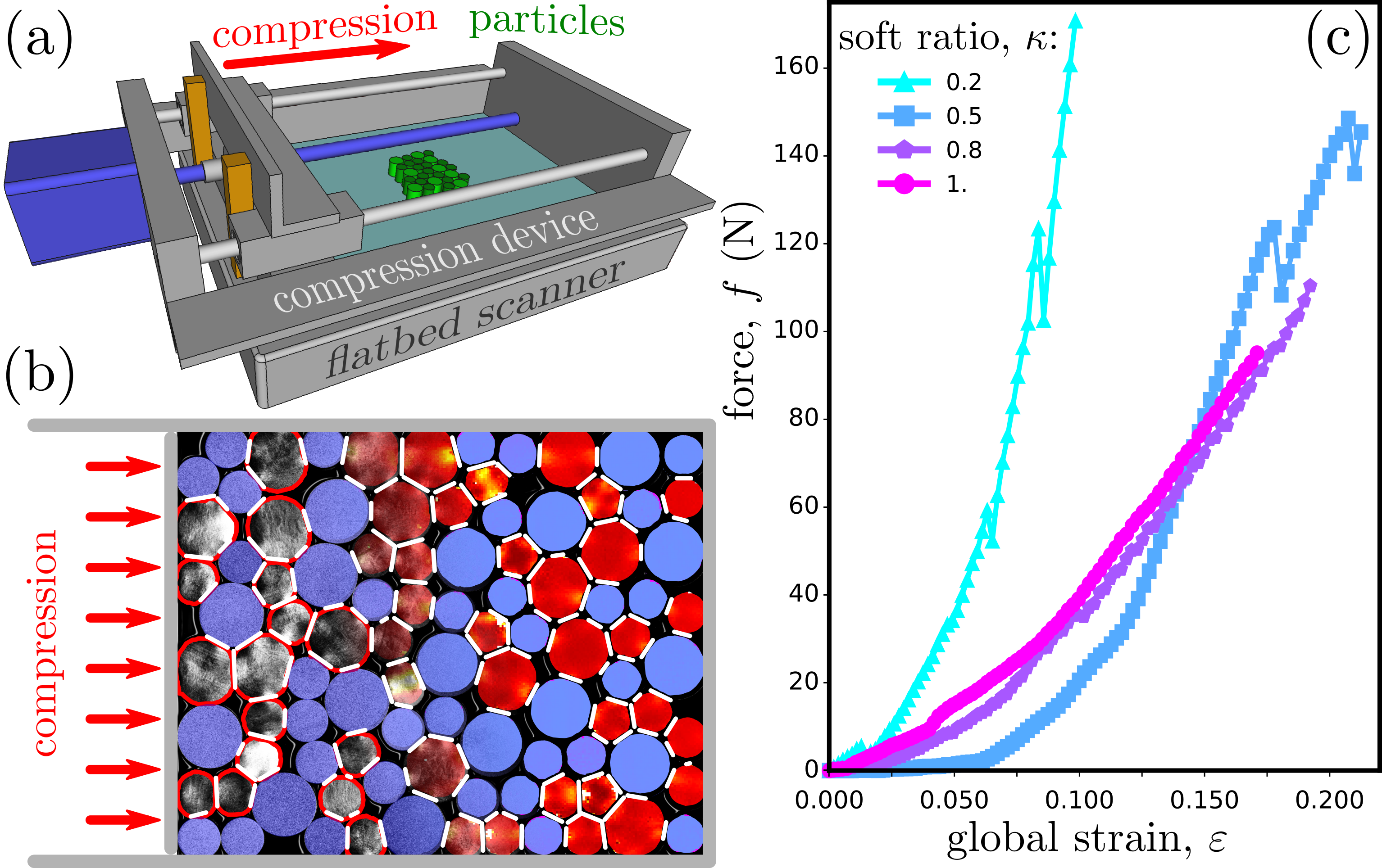}
\caption{(color online) a: Experimental set-up. A bidimensional bidisperse granular system, composed of soft and rigid particles, lays on the top glass of a flatbed scanner. A uniaxial compression device stresses it stepwisely while it is imaged from below by the scanner. b: Composite view of measured fields. Rigid particles are colored in blue. Raw image, in gray level, is shown on the left, particle boundaries are in red. Von Mises strain field, $\mathcal{C}$, is shown on the right with a color scale going from dark red (low value) to yellow (high value). Contacts are shown in white. c: Evolution of the measured global force, $f$, as a function of the global strain, $\varepsilon$, for different softness ratios, $\kappa$, given in the inset.}
\label{fig_1}
\end{figure}

\textit{Experiment} -- The experiments were carried out using a set-up partially introduced by \cite{vu2019_pre,bares2021_arx}. As shown in fig.~\ref{fig_1}a, it consists of a bidimensional piston of initial dimension $270 \times 202$~mm$^2$ which compresses uniaxially a bidisperse collection of rigid and soft cylinders laying on top of a flatbed scanner. The loading piston is composed of a stepper motor rotating a screw which translates the moving edge of the piston in the inward direction. Two force sensors are attached to this latter. They record the global force ($f$) evolution (see fig.~\ref{fig_1}c) when compressing the granular media. The induced global stress ($P$) is directly deduced and measured continuously at a frequency of $100$~Hz, while the system is compressed stepwisely. For each loading step the piston moves of $0.5$~mm at a speed of $2$~mm/min to stay in the quasistatic regime. Then, the system relaxes during $1$~min and is imaged from below with the scanner \cite{scanner} at a resolution of $2400$ dpi ($10.6$ $\mu$m/px).

The compressed granular systems are composed of both hyperelastic silicone cylinders \cite{silicone,vu2019_pre_2} and rigid PVC cylinders. Their Young moduli are $E_0=0.45$~MPa \cite{vu2019_em} and $1.2$~GPa, respectively and the Poisson ratio of silicone is $\nu=0.5$. Both soft and rigid granular packings are bidisperse cylinders of diameters $20$~mm and $30$~mm, and height $15$~mm. For each experiment, about $n=100$ particles are used, the ratio between rigid and small ones is the \textit {softness ratio}, $\kappa = \rm{n}_{\rm{soft}}/\rm{n}$. This ratio is equally spread among soft and rigid particles, and is varied from $0.2$ to $1$. To avoid basal friction the scanner glass is coated with oil \cite{bares2021_arx}. The bottom of each particle is coated with thin metallic glitter, which creates a random pattern with a correlation length of about $50$ $\mu$m \cite{vu2019_pre}.

For each experiment a set of about $90$ pictures of $\sim 500$~Mpx displays the evolution of the bottom face of the granular system. These pictures are post-processed with an algorithm modified from \cite{vu2019_pre, DIC_code}. First, particles are detected individually by thresholding the undeformed picture. Each particle is then tracked along the full set of pictures. Then, from the particle's solid rigid motion measurement, subsets of images are extracted following each particle and correcting its translation and rotation. For soft particles a DIC algorithm aimed for large deformations already presented in \cite{vu2019_em} is used to obtain the displacement field ($\vec{u}$) inside each particle. For the rigid particles a more classical DIC algorithm is used. This latter correlates all images with the initial one as classically done in the small deformation assumption. 

As shown in fig.~\ref{fig_1}b, the particle boundaries are obtained from the displacement fields $\vec{u}$ \cite{vu2019_pre}. The system packing fraction is directly deduced from it as well as the particle aspherisities, $a=p^2/(4 \pi s)$, where $p$ and $s$ are the perimeter and surface of the particles, respectively. For the soft particles, the right Cauchy-Green strain tensor field, $\vec{\vec{C}}$ is computed \cite{vu2019_pre,taber2004_bk} and its von Mises measure, $\mathcal{C}$, is deduced. For the rigid particles the local stresses are too low to induce any significant deformation of the PVC \cite{bares2021_arx}. Contacts between particles and their length ($l$) are measured from the proximity of the boundaries and the von Mises strain \cite{vu2019_pre}. The jamming transition \cite{liu1998_nat} at packing fraction $\phi_J$ is detected for each system for $P=1$~kPa and correspond with $Z_J=3.77 \pm 0.05$ (see SM) in agreement with the slipperiness of the particles \cite{van2009_jpcm}. 


\begin{figure}
\centering
\includegraphics[width=0.85\linewidth]{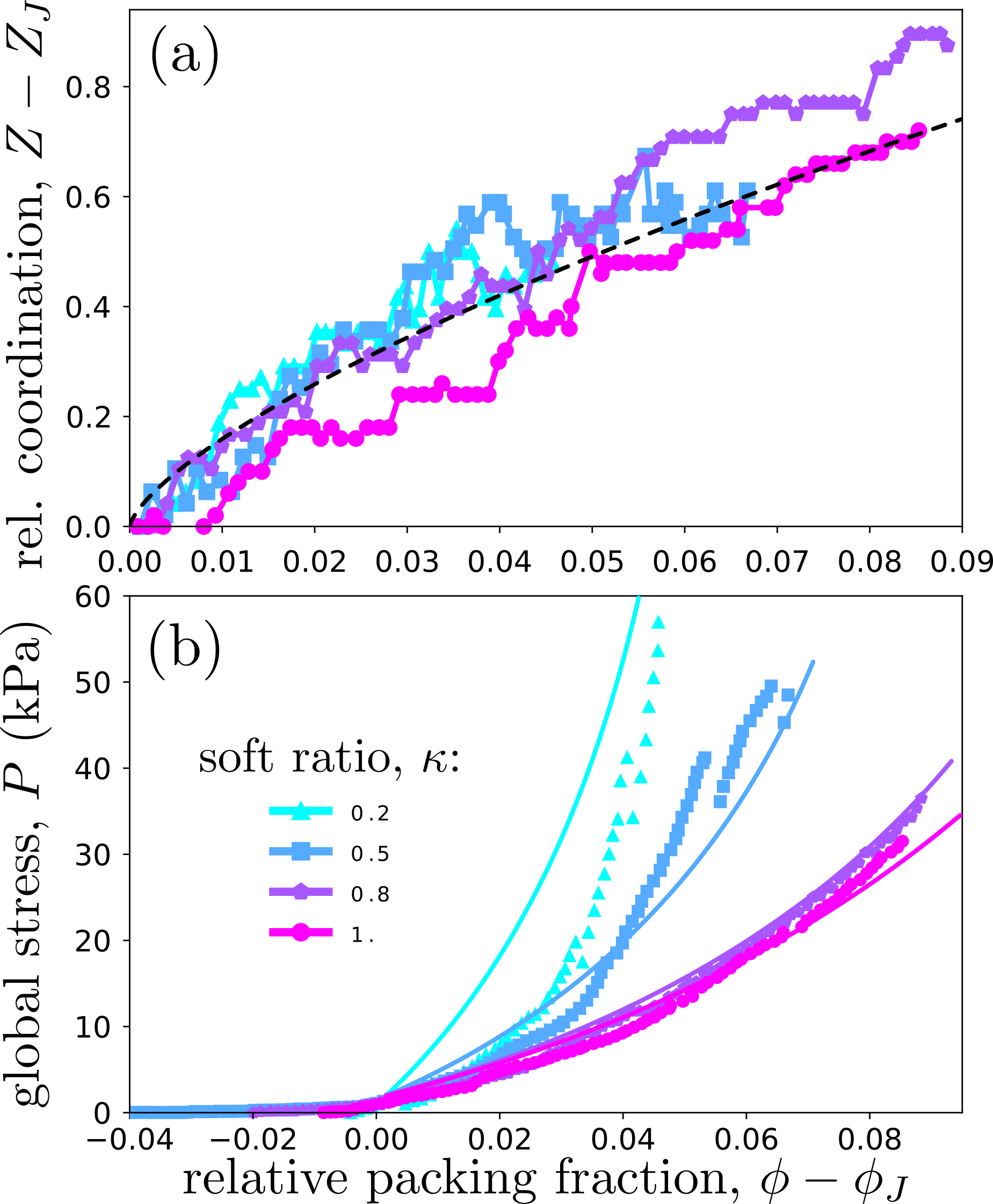}
\caption{(color online) a: Evolution of the coordination, $Z$, relative to the coordination at jamming, $Z_J \approx 3.8$, as a function of the distance to the jamming transition for the packing fraction, $\phi-\phi_J$. The dashed line corresponds to a power-law with exponent $0.7$. b: Evolution of the global stress applied to the system, $P$, as a function of $\phi-\phi_J$. Plane line curves correspond with eq.\ref{eq_1}. In both panels, the different curves correspond with different softness ratios, $\kappa$, given in inset of (b).}
\label{fig_2}
\end{figure}

\textit{Results} -- In the fig.~\ref{fig_2}a, we show how the increase of the coordination from the jamming point, $Z-Z_J$, evolves with the relative packing fraction, $\phi-\phi_J$. Whatever the softness ratio, curves collapse on a single one fitted by a power-law with exponent $\alpha = 0.7 \pm 0.15$ and prefactor $k = 4.8$. Within the error-bar, this value is close from what has been already observed \cite{andreotti2013_bk,nezamabadi2019_cpc,vu2019_pre_2}. 

Following a recent micro-mechanical compaction model numerically tested on isotropically compressed soft grains \cite{cantor2020_prl,cardenas2020_pre,cardenas2020_pre}, we extended it to the case of uniaxial compression (See supplemental material). In fig.~\ref{fig_2}b, compaction curves ($P$ \textit{vs.} $\phi-\phi_J$) are plotted with :
\begin{equation}
	P \simeq \frac{E^*}{4\Gamma} (1+\mu_M)  \frac{\phi_{\rm{max}}-\phi_J}{\phi_J}(Z_0+k(\phi-\phi_J)^{0.7})\phi\ln\left(\frac{\phi_{\rm{max}}-\phi_0}{\phi_{\rm{max}}-\phi}\right)
\label{eq_1}
\end{equation}
\noindent where $E^*=E_0/(2\kappa(1-\nu^2))$ is the mean effective material Young modulus and $\phi_{\rm{max}}$ is the asymptotic maximum packing fraction. For $\kappa=1$, $\phi_{\rm{max}}=0.97$ is close to $1$ while for $\kappa=0.2$, $\phi_{\rm{max}}=0.88$ closer to the random close packing (See supplemental material). $\Gamma$ is a geometrical parameter directly measured on the local vs. global strain curve (See supplemental material). $\mu_M$ is the macroscopic friction known to be close to 0.25 for an assembly of 2D particles \cite{estrada2008_pre,skinner1969_geo}. Although not based on fitting parameters, the prediction given by this model is excellent for all $\kappa$-values, we only note small disagreements with experimental data arising when the grains sharply rearrange. This effect is also exacerbated by the fact that the system is relatively small.


\begin{figure}
\centering
\includegraphics[width=0.85\linewidth]{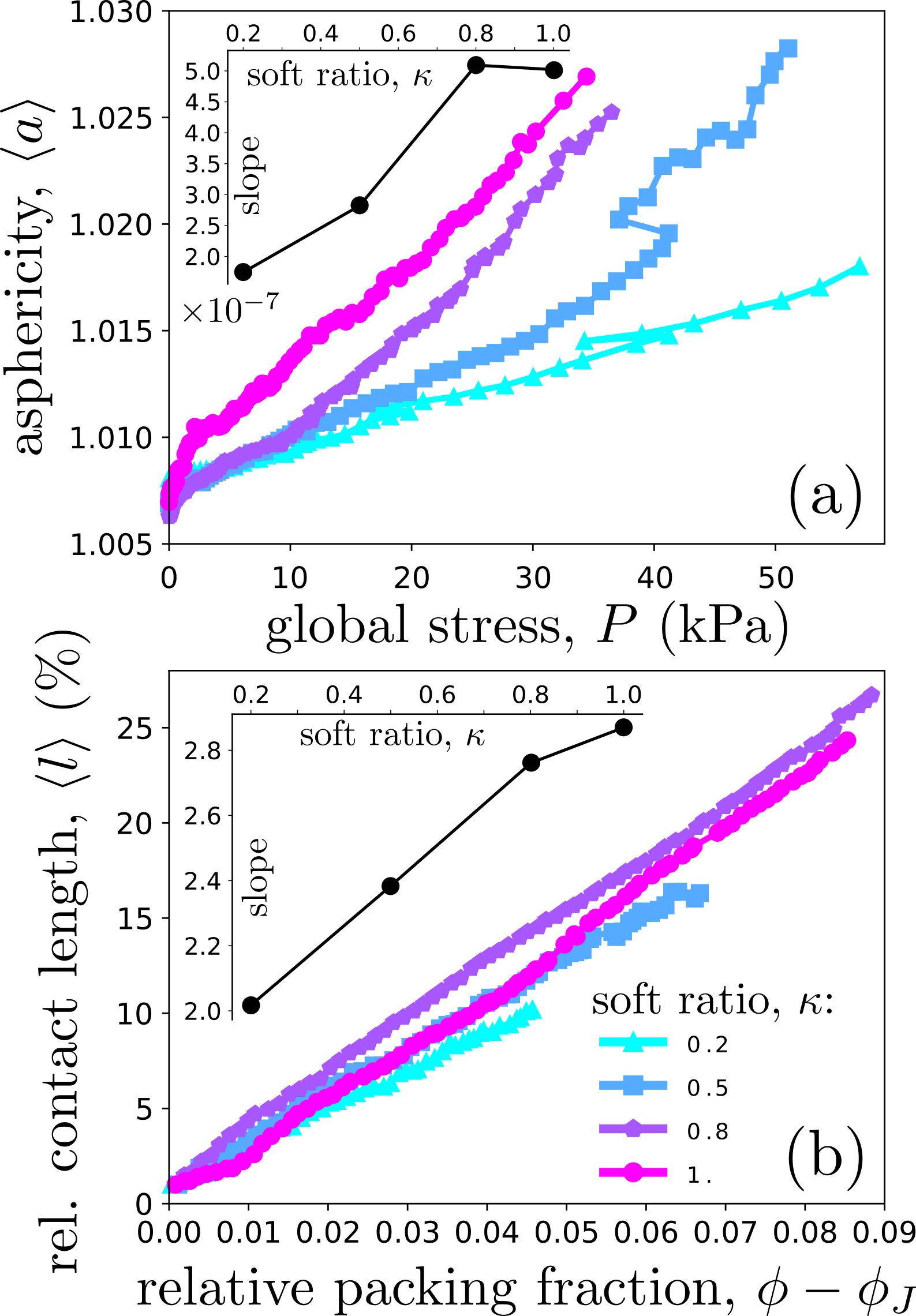}
\caption{(color online) a: Linear evolution of the soft particle average asphericity, $\left<a\right>$, as a function of the global stress applied to the system, $P$. Their slope is given in the inset as a function of the softness ratio, $kappa$. b: Linear evolution of the average value of the relative contact length, $\left<l\right>$, as a function of the distance to the jamming transition for the packing fraction, $\phi-\phi_J$. Their slope is given in the inset as a function of $\kappa$. In both panels the different curves correspond with different $\kappa$, given in inset of (b).}
\label{fig_3}
\end{figure}

Fig.~\ref{fig_3}a shows how the average soft particles asphericity, $\left<a\right>$, increases with global stress, $P$. Apart from few sharp grain rearrangements, this increase is linear whatever the softness ratio, $\kappa$. The slope increases linearly with $\kappa$ up to $0.8$ and then plateaus around $5 \times 10^{-7}$. This means the fewer particles are soft, the more they are deformed up to a point where this deformation saturates. In Fig.~\ref{fig_3}b we show the evolution of the average relative contact length, $\left<l\right>$. It increases linearly with the packing fraction for any $\kappa$ and the corresponding slope also increases linearly with this softness ratio before saturating around $2.8$ from $\kappa \approx 0.8$.


\begin{figure}
\centering
\includegraphics[width=0.85\linewidth]{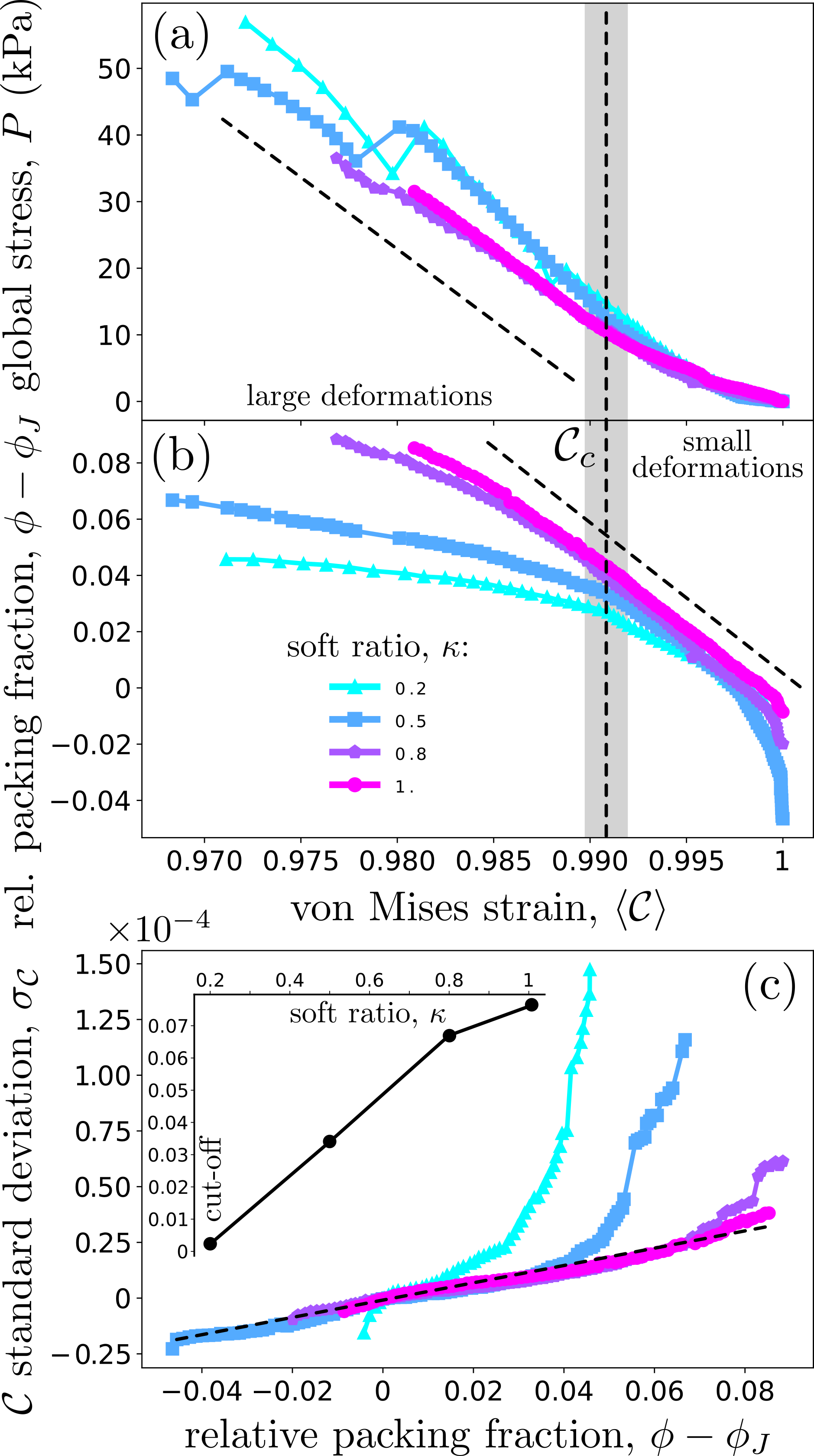}
\caption{(color online) a-b: Evolution of the global stress, $P$, and of the relative packing fraction $\phi - \phi_J$, as a function of the average value of the von Mises strain, $\left<\mathcal{C}\right>$ in soft particles, for different $\kappa$ values, given in inset of (b) (for the three panels). In panels (a) and (b) straight black dashed lines show a slope of $2.2$~kPa and $5.4$ respectively as a guide for the eyes. The vertical gray line and the shaded area show the cross-over strain $\mathcal{C}_c=0.991  \pm 1.2 \times 10^{-3}$ and its errorbar, respectively. It splits the space horizontally between small and large deformation levels. c: Evolution of the standard deviation of the von Mises strain in soft particles as a function of the relative packing fraction $\phi-\phi_J$. The straight dashed line with slope $4.0 \times 10^{-4}$ is a fit of the curves collapsed in their linear regime. These curves leave their linear regime at certain cut-off values given in the inset as a function of the softness ratio, $\kappa$.}
\label{fig_4}
\end{figure}

Under the small deformation assumption the infinitesimal strain tensor, $\vec{\vec{\varepsilon}}$, is related to $\vec{\vec{C}}$ following: $\vec{\vec{\varepsilon}} = 1/2 (\vec{\vec{C}} - \vec{\vec{I}})$, with $\vec{\vec{I}}$ the second order identity tensor. Hence, as shown in fig.~\ref{fig_4}a and b, when the system is compressed, $\mathcal{C}$ decreases from $1$ while the pressure and the packing fraction increase. At low compression level, for $\left<\mathcal{C}\right> > \mathcal{C}_c \approx 0.991$, the average von Mises strain, $\left<\mathcal{C}\right>$, decreases linearly with the packing fraction. In this regime, whatever the softness ratio, curves collapse fairly well. On the contrary at high compression level, for $\left<\mathcal{C}\right> < \mathcal{C}_c$, it decreases linearly with the global pressure. Not only the average strain evolves during compression, but its distribution also gets wider \cite{vu2019_pre} (See supplemental material). In fig.~\ref{fig_4}c, we show how the standard deviation of $\mathcal{C}$ increases with the packing fraction. When shifted from $\phi_J$, curves collapse on a linear law whatever $\kappa$, up to a cut-off which increases linearly with $\kappa$ up to a saturation point near $\kappa=0.8$.


\noindent\emph{Discussion \& conclusion} -- First, we point out that, even if the system seems small, macroscopic measurements stay repeatable. For example, the coordination at the jamming point is detected in a narrow band of $2.5\%$ for all the experiments. The exponent $\alpha$ of the $Z-Z_J$ \textit{vs.} $\phi-\phi_J$ power-law is slightly overestimated compared to previous experimental and numerical studies \cite{andreotti2013_bk,vu2019_pre_2,cantor2020_prl,cardenas2020_pre,cardenas2021_pre}. Still, considering the large error bars, it stays in the same domain: $\alpha \approx 0.5$. As explained in \cite{cantor2020_prl,cardenas2020_pre} this law can be plugged to existing models \cite{kim1987_ijp} to obtain an isotropic compaction equation. In the specific case of grain mixture, $\kappa$ is introduced in the effective Young Modulus to consider the softness variation. Also, to consider the uniaxiality of the compression, the induced effective friction is added in the law prefactor to obtain eq.\ref{eq_1}, and describe the compaction curve with no fitting parameters. In this latter equation, $\phi_{\rm{max}}$ goes ideally from $1$ for purely soft systems to $\phi_J$ for purely rigid systems, which is what we observe in our experimental data. So it is here remarkable that this model matches very well the experimental data whatever the grain mixture. This constitutes a validation of the model.

On top of extending and validating this compaction model, we introduce new relations between global observables and deformed local structures. Hence linear relations are observed between the particle asphericity and the global pressure ($\left<a\right> \sim P$) as well as for the relative contact length and the packing fraction ($\left<l\right> \sim (\phi-\phi_J)$). This first observation permits indirect measurement of pressure in a granular system just by measuring the grain boundary deformations. For this latter linear relation, in the small deformation regime, a scaling consideration of the Hertz contact law as well as $Z$ \textit{vs.} $\phi-\phi_0$ equation permits to predict this linear relation (See supplemental material). It is important to note that this linearity persists far beyond the small deformation regime and even for any particle mix, which makes it useful to indirectly deduce the packing fraction evolution of a compressed system from the contact geometry observation. In both linear relations, it is also remarkable that linear coefficients increase linearly with the softness ratio up to a point where it saturates near $\kappa=0.8$. This saturation comes from the fact that when $\kappa$ gets higher, for a given load or a given packing fraction, most of the deformation is carried out by a smaller number of grains which highly deform, leaving more rapidly the small deformation assumption and entering in a different, saturated, regime.

So far, never observed relations between global observables and the local deformation field are also revealed in our experimental results. For a small level of deformation, the average strain inside particles evolves linearly with the packing fraction ($\left<\mathcal{C}\right> \sim \phi-\phi_J$) while in the case of an important loading, it scales with the global pressure applied to the system ($\left<\mathcal{C}\right> \sim P$). This evidences two distinct regimes deep in the jammed state as already observed in \cite{vu2019_pre} and separated by a cross-over value $\mathcal{C}_c \approx 0.991$ independent of $\kappa$. In the first regime, linearity is reminiscent of the fact that the global strain scale with the packing fraction (See supplemental material), so this seems to be extrapolated to the local strain. In the second regime, linearity between strain and stress suggests that the material behaves like a bulk one (except for small rearrangements) this is consistent with the fact that the material is very dense with almost no interstitial porosity. It is also consistent with the fact that, at large compression levels, the pressure seems to decrease faster with the strain for lower softness ratio, the material gets stiffer for lower $\kappa$. On the contrary, linear relations collapse in the case of small deformations and do not significantly depend on the softness ratio.  
Finally, a scaling between strain standard deviation and packing fraction is evidenced ($\sigma_{\mathcal{C}} \sim \phi-\phi_J$) and does not depend on the softness ratio. Only the linearity cut-off increases with $kappa$. This is explained by the fact that for low $\kappa$ soft particles deform more and more rapidly, entering faster in a non-linear regime.

Granular systems composed of purely soft or mixed rigidity particles are ubiquitous, so we believe that the results obtained in this letter, more particularly the compaction law and scaling relations, can release blockages in domains as different as biology, geoscience or in the industry. Using our experimental set-up, these results could be confronted in the case of non circular particles or the case mixture of particles with different rheologies.



\begin{acknowledgments}

We thank Gilles Camp and St\'{e}phan Devic for their technical help with the experimental set-up and the particle making process.

\end{acknowledgments}


\bibliography{biblio.bib}

\end{document}